\documentclass{ws-p8-50x6-00}

\begin{document}

\title{Gamma Ray Bursts and vacuum polarization process in electromagnetic Black Holes}

\author{Remo J. Ruffini} 

\address{Physics Department 
  and \\
  International Center for Relativistic Astrophysics, %\\
  University of Rome, I--00185 Roma, Italy}

\maketitle

\abstracts{The developments of the elctromgnetic black holes physics and vacuum polarization process are presented in the interpretation of gamma ray bursts.}

\section{Introduction}
I have accepted with great pleasure the invitation to speak on our theory of Gamma ray Bursts by Pisin Chen for the interest of the program of this meeting which presents important developments on vacuum polarization effects in high energy collisions on which he himself has given so many important contributions.An additional factor has also been the beauty of the location of the school, in the Capri Island. What to me is also very important is that I follow the talks of two pioneers who have opened up some of the most exciting fields of Relativistic Astrophysics: I. F. Mirabel who has first pointed  the existence of so called supraluminal behaviour  in microquasars galactic sources and Livio Scarsi,  the ideator and leader of the Beppo Sax Satellite  which has discovered the afterglow of Gamma Ray Bursts sources. This last result has led to an unprecedented collaboration between all fields of space based optical, X and Gamma rays observatories in coincidence with the largest earth based observatories leading to the clear conclusion that Gamma ray Bursts are indeed of cosmological origin and their energy can be in some cases $\ge 10^{54}$ ergs (see Costa 2001\cite{c01}). An energy of $10^{54}$ ergs/burst can be easily visualized in its enormity: if we consider the light emitted by a star like our own sun and we multiply such luminosity by the number of stars in our galaxy ($10^{12}$) and multiply again such numbers by the total number of galaxies in the Universe ($10^9$), this is of the order of $10^{54}$ ergs/sec. In other words, during the burst the luminosity of a single GRB can  equal the  energy emission of the entire Universe. It exist the very clear possibility that the interpretation of both these phenomena, the Micrroquasars and the GRBs, are deeply rooted in the physics of Black Holes to which I have devoted special attention in my theoretical research. I will give some motivation for this possibility in the following.

\section{On three happenings related to GRBs}

The case of Gamma Ray Bursts is for me personally very intriguing. There have been moments in my life which appear to be intertwined with some of the relevant events that are leading to the understanding of such most unique phenomena. Moreover each  scientific contribution I have achieved and even apparently occasional occurrences in my life seemingly disconnected appear to acquire special meaning in reaching  the understanding of such phenomena. 

The first of such happening was the collaboration during my career, started at Princeton  in 1967 with John Archibald Wheeler and the one started in 1968 at Tbilisi and then in Moscow with Yakov Borisovich Zel'dovich. Both these scientists were heads and founders of the research in relativistic astrophysics in their countries. But also interesting was the fact that both of them had a principal role, before adressing their attention to the implications of Einstein theory of general relativity to astrophysics, in the nuclear arms race respectively in the USA and Soviet Union. Johnny Wheeler was a leader in the first tactic H Bomb explosdion on the Bikini Atoll.  Ya. B. Zeldovich  after having contributed to the defense of his country with the Katiuscia rockets had developed with Andrej Sakahrov, the soviet A and H bomb. He was also the proposer of a most unusual and intolerable project to have an H bomb explode on the far side of the moon to demonstrate at ounce the ``maturity'' of the nuclear and space technology reached by the Soviet Union in the early sixties. 

\begin{figure}
\vspace{-.5cm}
\epsfxsize=6.0cm
\begin{center}
\mbox{\epsfbox{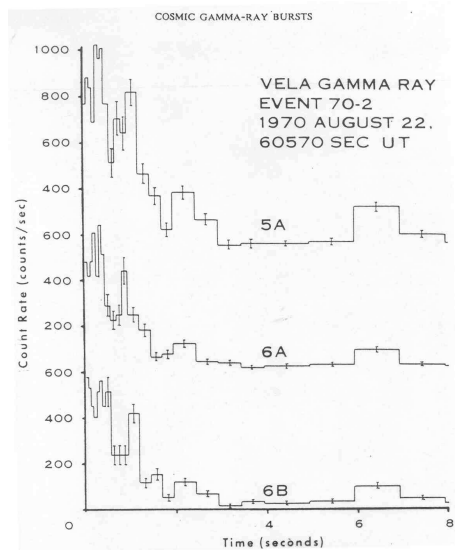}}
\end{center}
\vspace{-0.2cm}
\caption[]{One of the first GRBs observed by the Vela satellite. Reproduced from Strong in Gursky \& Ruffini (1975)\cite{gr75}.}
\label{velaburst}
\end{figure}

The second happening occurred in 1975. Herbert Gursky and myself had been invited by the AAAS to organize a session on neutron stars, black holes and binary X ray sources for their annual meeting in San Francisco. During the preparation of the meeting we heard that some observations made by the military Vela satellites, conceived in order to monitor the Limited Test Ban Treaty of 1963 banning atomic bomb explosions, had just been unclassified. Doubtlessly the  unorthodox proposal of Zel'dovich  had been among the motivations to develop such a grandiose militar monitoring system. We asked Ian B. Strong to report, for the first time in a public meeting, on these just observed-released gamma ray bursts (GRBs)(Strong 1975)\cite{gr75}, See Fig.~\ref{velaburst}.

\begin{figure}
\vspace{-.5cm}
\epsfxsize=\hsize
\begin{center}
\mbox{\epsfbox{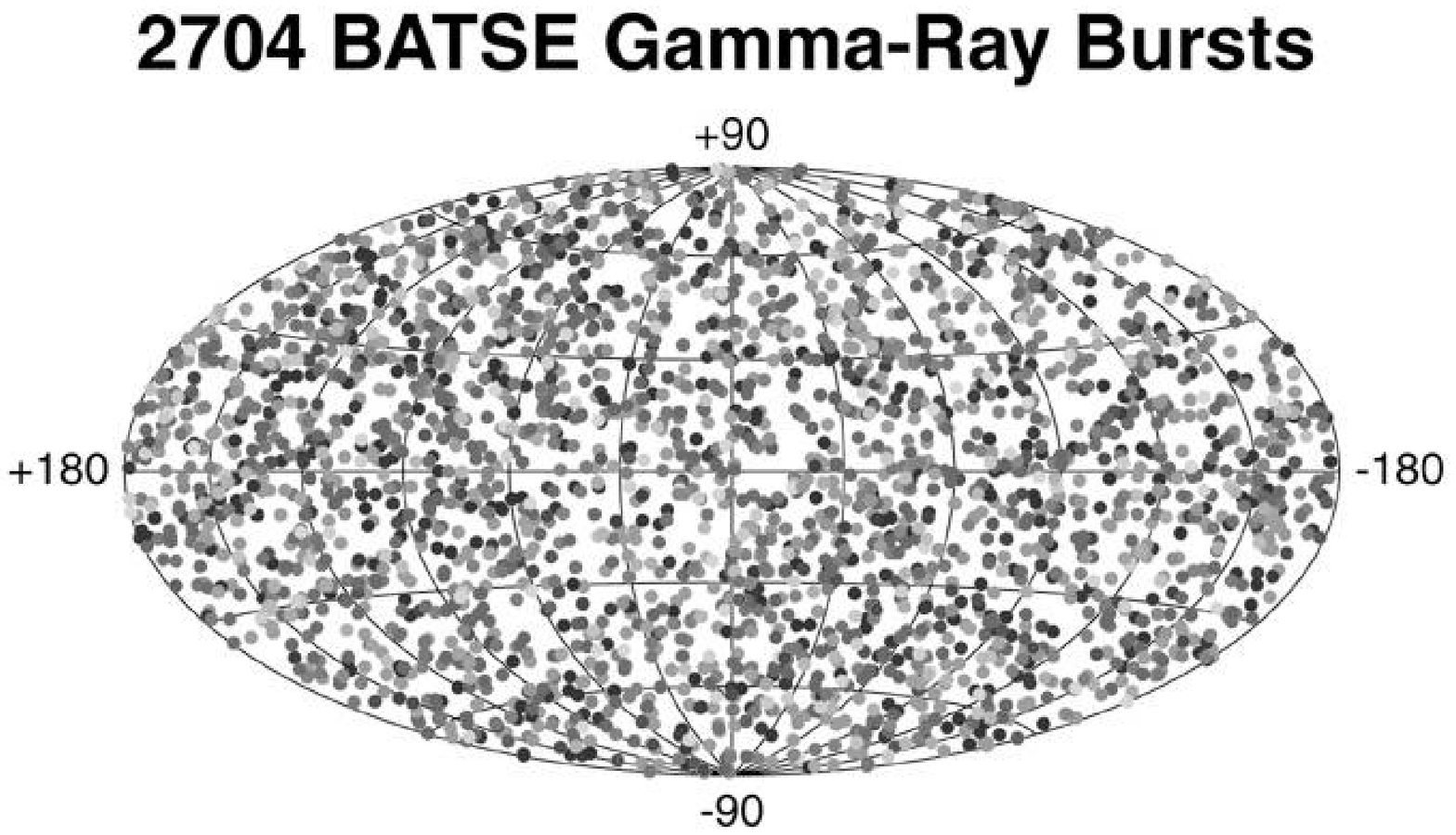}}
\end{center}
\vspace{-0.2cm}
\caption[]{Angular distribution of GRBs in galactic coordinates from the Compton GRO satellite.}
\label{batsedist}
\end{figure}

It was clear since the earliest observations that these signals were not coming either from the Earth or the planetary system. By 1991 a great improvement in knowledge of the distribution of the GRBs came with the NASA launch of the Compton Gamma-Ray Observatory which in ten years of observations gave beautiful evidence for the perfect isotropy of the angular distribution of the GRB sources in the sky, see Fig.~\ref{batsedist}. The sources had to either be at cosmological distances or very close to the solar system in order not to reflect the anisotropic galactic  distribution.

The third happening occurred in 1989 when I was elected President of the scientific committee of the Italian Space agency (ASI) and the committee found itself involved with the scrutiny of the first Italian scientific satellite: the SAX satellite. The project was a collaboration between Italy and Netherland: The total estimated cost was roughly 50 millions US dollars, fairly shared by the two partners: 25 milions for Italy and 25 for the Netherlands. The satellite was supposed to fly in 1985. The program had already been delayed four years by the time of our scrutiny started. The costs had correspondingly ``skyrocketed'' to almost 250 millions US dollars,"fairly" shared by the partners: 225 millions from Italy and 25 from the Netherlands... The real moment of panic came when we learned that the Dutch had run out of money. They could not afford to pay for the wide field x-ray cameras, they were supposed to contribute. We decided to intervene offering to pay roughly six millions US dollars from the limited budget of our committee in order to avoid any further delay and especially to avoid the loss  of one of the crucial instruments of the scientific mission. As the delays were augmenting and the expenses of the mission were correspondingly  ``skyrocketing'' further,  the ambiance soon deteriorated to inadmissible pressures.Before quitting I insisted on the imperative to accomplish the mission no matter what. In these hard days Livio  and I were on opposite sides, but we managed to keep the highest personal and scientific esteem for each other. 

The satellite finally was launched in 1996 at a cost still today unknown: it  is not yet possible to ascertain  if it passed and, if so, of how much the one billion Us dollars mark. Soon after the lunch three of the four gyroscopes failed apparently due to the improper choice of space qualified instruments and the satellite was  left without an effective pointing capability.

 In spite of all that, thanks also to the determinate action of a number of strongly dedicated and courageous young physicists educated at ``La Sapienza'' who had joined the Milano based original  team, the newly named Beppo-SAX  satellite  was able to conclude one of the most successful  ever  scientific missions in Astronomy and Astrophysics. They discovered the afterglows of the GRBs, which in turn have allowed the optical identification of the sources and the  determination of their cosmological nature. I am happy  today to see Livio here again in the presentation of these most beautiful scientific results.  I am also happy to see that  also the completion of the wide field x ray camera and the many imperatives to conclude the mission have lead to a successful  epilogue. 

Thinking about this past situation with hinsight, I have reached a rather unorthodox conclusions: if SAX had flown on time in 1985 and possibly within its planned budget, it would have been a managerial success and quite a saving for the Italian treasury, but very likely not a scientific success. The reason is that in 1985  neither the Space Telescope nor the very large telescopes like KECK  and  VLT, which have been essential to the optical identification of the GRBs and the establishment of their cosmological distance, were functioning. Of course I do not want to make propaganda in favour of wrong doings, but it appears as if a tremendous force directs human actions not only exploiting great scientific ideas but making use as well of weakness, mistakes and mismanagement in order to reach a final important scientific goal! 

\begin{figure}
\vspace{-.5cm}
\epsfysize=17.0cm
\begin{center}
\mbox{\epsfbox{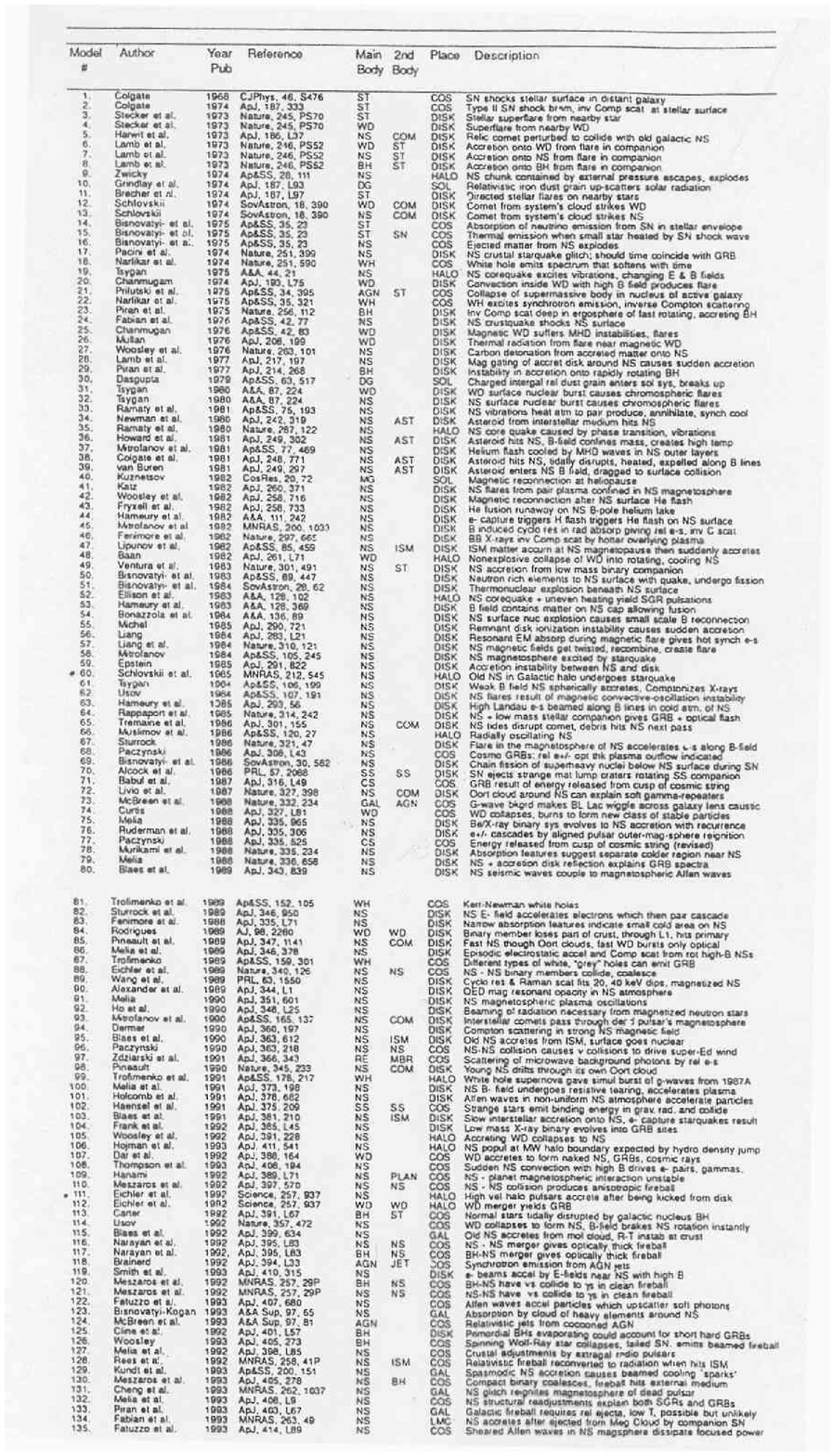}}
\end{center}
\vspace{-0.2cm}
\caption[]{Partial list of theories before the Beppo-SAX, from a talk presented at MGIXMM\cite{mgixmm}.}
\label{teofig}
\end{figure}

	While the expenditure on Beppo-SAX were ``skyrocketing'', equally exponentially increasing were the numbers of competing theories trying  to explain GRBs. see a partial list in Fig.~\ref{teofig} 

The observations of the Beppo-SAX satellite had a very sobering effect on the theoretical developments for GRB models.Almost the totality of the existing theories, see above partial list, were at once wiped out, not being able to fit the stringent energetics requirements imposed by the observations. 

This led us to the return to the model of GRBs, which we had quietly advanced and theoretically developed in all its conceptual complexity starting from the work with Thibau Damour in 1975\cite{dr75}, essentially based on the possibility of extracting mass-energy from a Black Hole. Such a process is based on the  mass-energy formula for Black Holes I had found in 1971 with Demetrios Christodoulou\cite{ruffc}. Following this work it had become evident that Black Holes far from being energy sinks were in principle the most energetic source of energy in the Universe. The extraction of energy from Black Holes can  in fact surpass, in principle, chemical, nuclear or thermonuclear energy source both in absolute value and intensity. In practice it is now clear that this happens in GRBs as predicted by our model.

 I will shortly recall  the background on which our model was conceived, how the concepts of ``alive''  versus ``dead'' black holes was introduced, the key role of the reversible and irreversible transformations in Black Holes and the role of  quantum vacuum polarization process, and  finally indicate some current developments.

\section{Early steps in the study of Black Holes}

The year 1968 with the discovery of pulsars in 1968 and especially with the discovery of the pulsar in the Crab Nebula, can be considered the birthdate of relativistic astrophysics. The observation of the period of that pulsar and its slow-down rate  not only clearly gave unequivocal evidence for the identification of the first neutron star in the galaxy but also contributed to the understanding that the energy source of pulsars is very simply the rotational energy of a neutron star.

\begin{figure}
\vspace{-.5cm}
\epsfxsize=6.0cm
\begin{center}
\mbox{\epsfbox{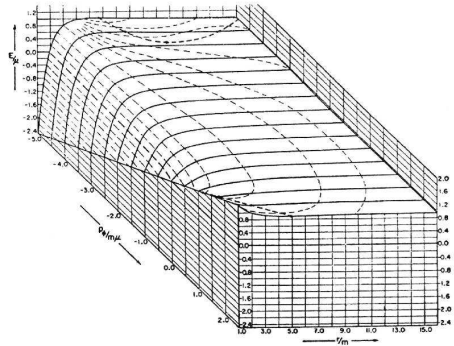}}
\end{center}
\vspace{-0.2cm}
\caption[]{``Effective potential'' around a Kerr black hole, see Ruffini and Wheeler 1971}
\label{poten}
\end{figure}

\begin{figure}
\vspace{-.5cm}
\epsfxsize=6.0cm
\begin{center}
\mbox{\epsfbox{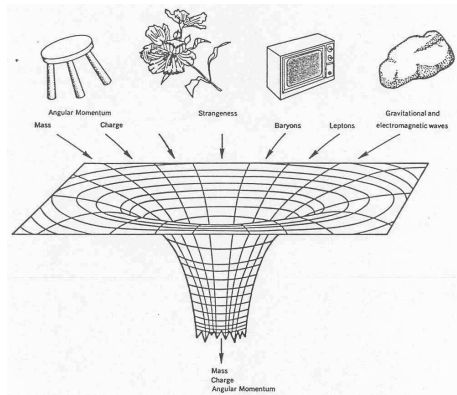}}
\end{center}
\vspace{-0.2cm}
\caption[]{The black hole uniqueness theorem.}
\label{tvsetbh}
\end{figure}

I was in Princeton in those days initially as a  postdoctoral fellow at the university in the group of John Archibald Wheeler, then as a member of the Institute for Advanced Study, and finally as an instructor and assistant professor at the University. The excitement over the neutron star discovery  boldly led us directly to an as yet unexplored classic paper by Robert Julius Oppenheimer and Snyder ``on continued gravitational contraction''\cite{snyder} and this opened up an entirely new field of research to which I have dedicated the remainder of my life and it is still producing some quite important results today. An ``effective potential'' technique had been used very successfully by Carl St\o rmer in the 1930s in studying the trajectories of cosmic rays in the Earth's magnetic field (St\o rmer 1934)\cite{s34}. In the fall of 1967 Brandon Carter visited Princeton and presented his remarkable mathematical work leading to the separability of the Hamilton-Jacobi equations for the trajectories of charged particles in the field of a Kerr-Newmann geometry (Carter 1968)\cite{bc}. This visit had a profound impact on our small group working with John Wheeler on the physics of gravitational collapse. Indeed it was Johnny who had the idea of exploiting the analogy between the trajectories of cosmic rays and spacetime trajectories in general relativity, using the St\o rmer ``effective potential'' technique in order to obtain physical consequences from the set of first order differential equations obtained by Carter. I still remember the excitement of preparing the $2m\times 2m$ grid plot of the effective potential for particles around a Kerr black hole which finally appeared later in print (Rees, Ruffini and Wheeler 1973,1974\cite{rrw}; see Fig.~(\ref{poten}). 
Out of this work came the celebrated result for the maximum binding energy  $1 - {1 \over \sqrt{3}}\sim42\%$ for corotating orbits and $1-{5\over 3\sqrt{3}}\sim 3.78\%$ for counter-rotating orbits in the Kerr geometry. We were very pleased to be later associated with Brandon Carter in a ``gold medal'' award for this work presented by Yevgeny Lifshitz: in the fourth and last edition of volume 2 of the Landau and Lifshitz series ({\it The Classical Theory of Fields\/} 1975), both Brandon's work and my own work with Wheeler were proposed as named exercises for bright students! In our article ``Introducing the Black Hole'' (Ruffini and Wheeler 1971)\cite{rw71} we first proposed the famous ``uniqueness theorem'' stating that black holes can only be characterized by their mass-energy $E$, charge $Q$ and angular momentum $L$. This analogy between a black hole and a very elementary physical system  was imaginatively represented by Johnny in a  very unconventional figure in which TV sets, bread, flowers and other objects lose their characteristic features and merge in the process of gravitational collapse into the three fundamental parameters of a black hole, see Fig.~\ref{tvsetbh}. That picture became the object of a great deal of lighthearted discussion in the physics community. A proof of this uniqueness theorem, satisfactory for some case of astrophysical interest, has been obtained after twenty five years of meticulous mathematical work (see e.g., Regge and Wheeler\cite{ReggeW}, Zerilli\cite{Zerilli1,Zerilli2}, Teukolsky\cite{teukolsky}, C.H. Lee\cite{lee}, Chandrasekhar\cite{chandra}.) However, the proof  still presents some outstanding technical difficulties in its most general form. Possibly some progress will be reached in the near future with the help of computer algebraic manipulation techniques to overcome the extremely difficult mathematical calculations (see e.g., Cruciani (1999) {cru}, Cherubini and Ruffini  (2000)\cite{chrr} Bini et al.\ (2001)\cite{bcjr1}, Bini et al.\  (2001)\cite{bcjr2}).

It is interesting that this analogy, which appeared at first to be almost trivial, has revealed itself to be one of the most difficult to be proved requiring an enormous effort, unsurpassed in difficulty both in mathematical physics and relativistic field theory.

I am profoundly convinced that the solution of this problem from a mathematical physics point of view will have profound implications for our understanding of the fundamental laws of physics.

\section{ From ``dead'' to ``alive'' black Holes }

We were still under the sobering effects of the pulsar discovery and the very clear explanation by Tommy Gold and Arrigo  Finzi that the rotational energy of the neutron star had to be the energy source of the pulsar phenomenon, when the first meeting of the European Physical Society took place in Florence in 1969. In a stimulating talk Roger Penrose\cite{p69} advanced the possibility that, much like in the case of pulsars, the rotational energy of black holes could, in principle, be extracted in an analogous way.
\begin{figure}
\vspace{-.5cm}
\epsfxsize=6.0cm
\begin{center}
\mbox{\epsfbox{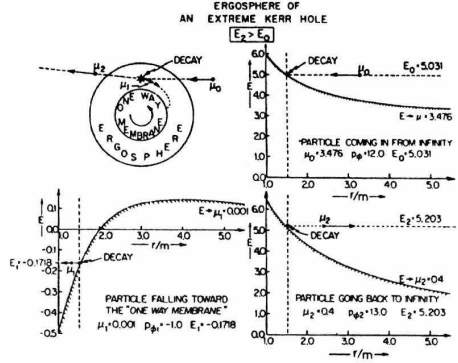}}
\end{center}
\vspace{-0.2cm}
\caption[]{Decay of a particle of rest-plus-kinetic energy $E_\circ$ into a particle which is captured by the black hole with positive energy as judged locally, but negative energy $E_1$ as judged from infinity, together with a particle of rest-plus-kinetic energy $E_2>E_\circ$ which escapes to infinity. The cross-hatched curves give the effective potential (gravitational plus centrifugal) defined by the solution $E$ of Eq.(2) for constant values of $p_\phi$ and $\mu$. (Figure and caption reproduced from Christodoulou 1970\cite{chris1}, in turn reproduced before its original publication in ref.~\cite{rw71} with the kind permission of Ruffini and Wheeler.)}
\label{pic1}
\end{figure}
The first specific example of such an energy extraction process by a gedanken experiment was given using the above-mentioned effective potential technique in Ruffini and Wheeler (1970)\cite{ruffx}, see Figure (\ref{pic1}), and then later by Floyd and Penrose (1971)\cite{fr}. The reason for showing this figure here is a) to recall the first explicit computation and b) to recall the introduction of the ``ergosphere'', the region between the horizon of a Kerr-Newmann metric and the surface of infinite redshift were the energy extraction process can occur, and also c) to emphasize how contrived, difficult and also conceptually novel such an energy-extraction mechanism can be. It is a phenomenon which is not localized at a point but which can occur in an entire region: a global effect which relies essentially on the concept of a field. It can only work, however, for very special parameters and is in general associated with a reduction of the rest mass of the particle involved in the process. It is almost trivial to slow down the rotation of a black hole and increase its horizon by  accretion of counter-rotating particles, but it is extremely difficult to extract the rotational energy from a black hole by a slow-down process, as also clearly pointed out by the example in Fig.~(\ref{pic1}). The establishment of this analogy offered us the opportunity to appreciate once more the profound difference between seemingly similar effects in general relativity and classical field theories. In addition to the existence of totally new phenomena, like the dragging of the inertial frames around a rotating black hole for example, we had the first glimpse of an entirely new field of theoretical physics present in and implied by the field equations of general relativity. The deep discussions of these problems with Demetrios Christodoulou, who was a 17 year old Princeton student at the time,  my first graduate student, led us to the discovery of  the existence in black holes physics of both ``reversible and irreversible transformations''. 

\begin{figure}
\vspace{-.5cm}
\epsfxsize=6.0cm
\begin{center}
\mbox{\epsfbox{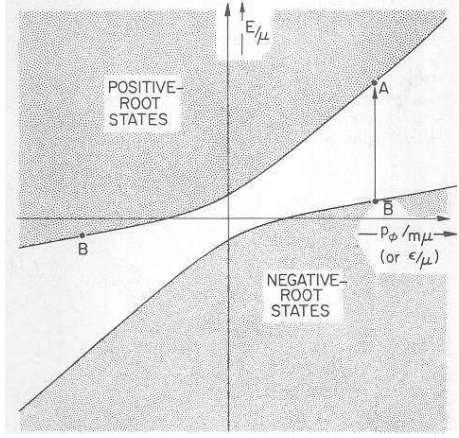}}
\end{center}
\vspace{-0.2cm}
\caption[]{Reversing the effect of having added to the black hole one particle (A) by adding another particle (B) of the same rest mass but opposite angular momentum and charge in a ``positive-root negative-energy state''. Addition of B is equivalent to subtraction of $B^-$. Thus the combined effect of the capture of particles A and B is an increase in the mass of the black hole given by the vector $B^-A$. This vector vanishes and reversibility is achieved when and only when the separation between positive root states and negative root states is zero, in this case the hyperbolas coalesce to a straight line. Reproduced from\cite{ruffc}.}
\label{posneg}
\end{figure}

Indeed it was by analyzing the capture of test particles by a Black Hole endowed with electromagnetic structure, for short an EMBH,  that we identified a set of limiting transformations which did not affect the surface area of an EMBH. These special transformations had to be performed very slowly, with a limiting value of zero kinetic energy on the horizon of the EMBH, see Fig.~\ref{posneg}. It became then immediately clear that the total energy of an EMBH could in principle be expressed as a ``rest energy'' a ``Coulomb energy'' and a ``rotational energy''. The rest energy is ``irreducible'', the other two being submitted to positive and negative variations, corresponding respectively to process of addition and extraction of energy.

While Wheeler was mainly studying the thermodynamical analogy, I addressed with Demetrios the fundamental issue of the energetics of EMBHs using the tools of reversible and irreversible transformations. We finally obtained the general mass-energy formula for black holes (Christodoulou and Ruffini 1971)\cite{ruffc}: 
\begin{eqnarray}
E^2&=&M^2c^4=\left(M_{\rm ir}c^2 + {Q^2\over2\rho_+}\right)^2+{L^2c^2\over \rho_+^2}\,,\label{em}\\
S&=& 4\pi \rho_+^2=4\pi \left(r_+^2+{L^2\over c^2M^2}\right)=16\pi\left({G^2\over c^4}\right) M^2_{\rm ir}\,,
\label{sa}
\end{eqnarray}
with
\begin{equation}
{1\over \rho_+^4}\left({G^2\over c^8}\right)\left( Q^4+4L^2c^2\right)\leq 1\,,
\label{s1}
\end{equation}
where $M_{\rm ir}$ is the irreducible mass, $r_{+}$ is the horizon radius, $\rho_+$ is the quasi-spheroidal cylindrical coordinate of the horizon evaluated at the equatorial plane, $S$ is the horizon surface area, and extreme black holes satisfy the equality in eq.~(\ref{s1}). The crucial point is that transformations at constant surface area of the black hole, namely reversible transformations, can release an energy up to 29\% of the mass-energy of an extremal rotating black hole and up to 50\% of the mass-energy of an extremely magnetized and charged black hole. Since my Les Houches lectures ``On the energetics of black holes'' (B.C. De Witt 1973)\cite{dw73} I introduced the concepts of ``alive'' black hoes, endowed of mas-energy ,of rotation and angular momentum and ``dead'' black holes uniquely characterized by their masses: one of my main research goals has been to identify an astrophysical setting where the extractable mass-energy of the black hole could manifest itself. As we will see in the following, I propose that this extractable energy of an EMBH is the energy source of gamma-ray bursts (GRBs).

\section{The paradigm for the identification of the first ``black hole'' in our galaxy and the development of X-ray astronomy.}

The launch of the ``Uhuru'' satellite by the group directed by Riccardo Giacconi at American Science and Engineering, dedicated to the first systematic examination of the universe in X-rays,  marked a fundamental leap forward and generated a tremendous momentum in the field of relativistic astrophysics.  The very fortunate collaboration soon established with simultaneous observations in the optical and radio wavelengths allowed generated high quality data on binary star systems composed of a normal star being stripped of matter by a compact massive companion star: either a neutron star or a black hole.

The ``maximum mass of a neutron star'' was the subject of the thesis of Clifford Rhoades, my second graduate student at Princeton. A criteria was found there to overcome fundamental unknowns about the behavior of matter at supranuclear densities by establishing an absolute upper limit to the neutron star mass based only on general relativity, causality and the behaviour of matter at nuclear and subnuclear densities.   This work, presented at the 1972 Les Houches Summer School (B. and C. de Witt 1973), only appeared after a prolongued debate (see the reception and publication dates!) (Rhoades and Ruffini 1974)\cite{rr74}.

The three essential components in establishing the paradigm for the identification of the first black hole in Cygnus X1 (Leach and Ruffini 1973)\cite{lr73} were
\begin{itemize}
\item the ``black hole uniqueness theorem'', implying the axial symmetry of the configuration and the absence of regular pulsations from black holes,
\item the ``effective potential technique'', determining the efficiency of the energy emission in the accretion process, and 
\item  the ``upper limit on the maximum mass of a neutron star'', discriminating between an unmagnetized neutron star and a black hole.
\end{itemize}
These results were also presented in a widely attended session chaired by John Wheeler at the 1972 Texas Symposium in New York, extensively reported on by the New York Times. The New York Academy of Sciences which hosted the symposium had just awarded me their prestigious Cressy Morrison Award for my work on neutron stars and black holes. Much to their dismay I never wrote the paper for the proceedings since it coincided with the one submitted for publication (Leach and Ruffini 1973)\cite{lr73}.

The definition of the paradigm did not come easily but slowly matured after innumerable discussions, mainly by phone, both with Riccardo Giacconi and Herb Gursky. I still remember an irate professor of the Physics Department at Princeton pointing out publicly my outrageous phone bill of \$274 for one month, at the time considered scandalous, due to my frequent calls to the Smithsonian, and a much more relaxed and sympathetic attitude about this situation held by the department chairman, Murph Goldberger. This work was finally summarized in two books: one with Herbert Gursky (Gursky and Ruffini 1975)\cite{gr75}, following the 1973 AAAS Annual Meeting in San Francisco, and the second with Riccardo Giacconi (Giacconi and Ruffini 1978)\cite{gr78} following the 1975 LXV Enrico Fermi Summer School (see also the proceedings of the 1973 Solvay Conference). 

\section{the Heisenberg-Euler critical capacitor and the vacuum polarization around a macroscopic black hole}

\begin{figure}
\vspace{-.5cm}
\epsfxsize=\hsize
\begin{center}
\mbox{\epsfbox{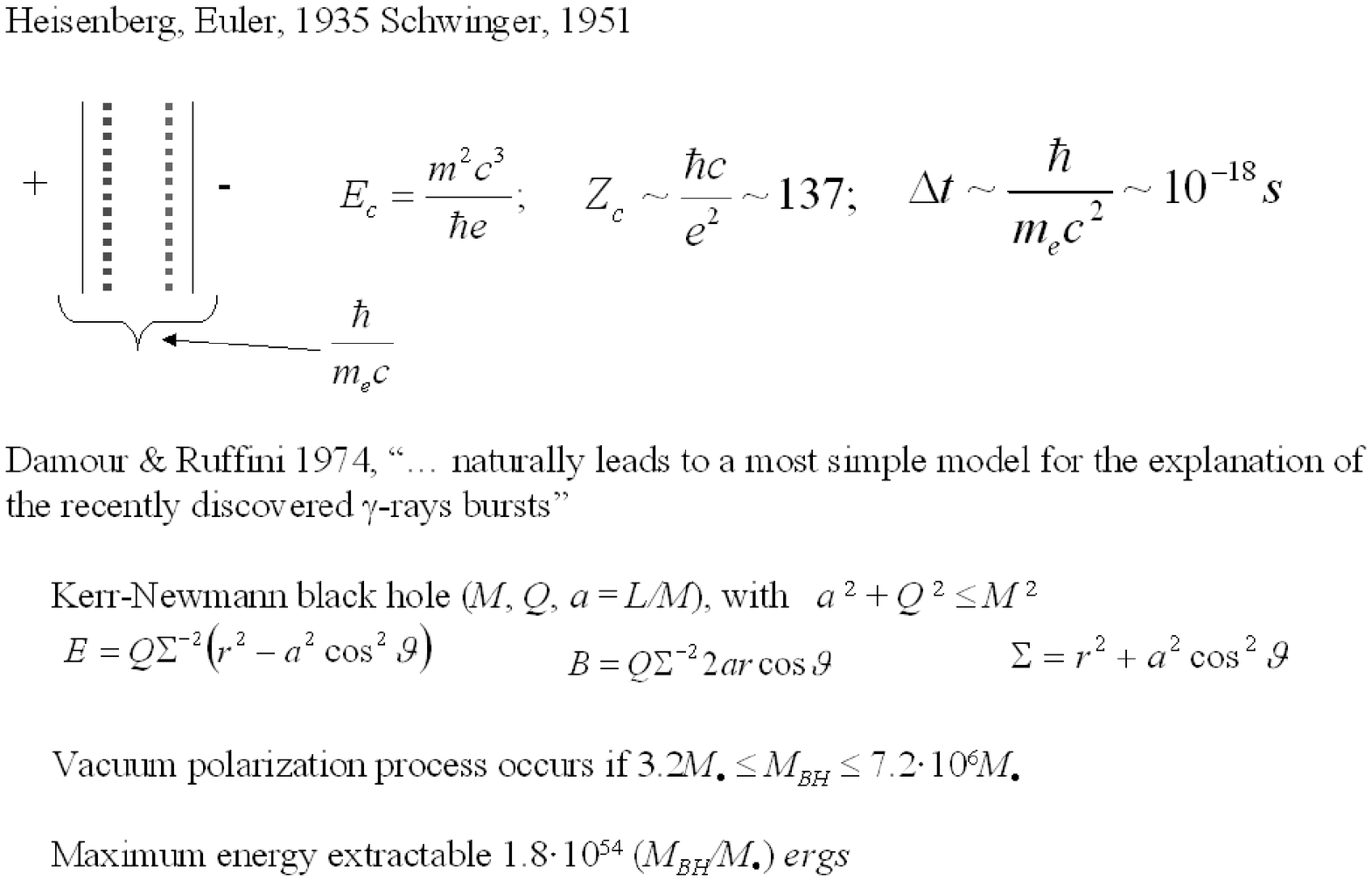}}
\end{center}
\vspace{-0.2cm}
\caption[]{Summary of the EMBH vacuum polarization process. See Damour \& Ruffini (1975)\cite{dr75} for details.}
\label{capdiap}
\end{figure}

In 1975, following the work on the energetics of black holes (Christodoulou and Ruffini 1971)\cite{ruffc}, we pointed out (Damour and Ruffini, 1975)\cite{dr75} the existence of the vacuum polarization process {\it a' la} Heisenberg-Euler-Schwinger (Heisenberg and Euler 1935\cite{he35}, Schwinger 1951\cite{s51}) around black holes endowed with electromagnetic structure (EMBHs). Such a process can only occur for EMBHs of mass smaller then $7.2\cdot 10^{6}M_\odot$. The basic energetics implications were contained in Table~1 of that paper (Damour and Ruffini, 1975)\cite{dr75}, where it was also shown that this process is almost reversible in the sense introduced by Christodoulou and Ruffini (1971)\cite{ruffc} and that it extracts the mass energy of an EMBH very efficiently. We also pointed out that this vacuum polarization process around an EMBH offered a natural mechanism for explaining GRBs, just discovered at the time,  and the characteristic energetics of the burst could be $\ge 10^{54}$ ergs, see Fig.~\ref{capdiap}.

\section{ the ergosphere versus the  dyadosphere of a black hole}

The enormous energy requirements of GRBs evidenced by the Beppo-SAX satellite, very similar to the ones predicted in Damour \& Ruffini (1975)\cite{dr75} have convinced us to return to our earlier work in studying more accurately the process of vacuum polarization and the region of pair creation around an EMBH. This has led to a) the new concept of the dyadosphere of an EMBH (named for the Greek word {\it dyad\/} for pair) and b) the concept of a plasma-electromagnetic (PEM) pulse and c) the analysis its temporal evolution generating signals with the characteristic features of a GRB.

In our theoretical approach, we claim that through the observations of GRBs, we are witnessing the formation of an EMBH and therefore follow the process of gravitational collapse in real time. Even more important, the tremendous energies involved in the energetics of these sources have their origin in the extractable energy of black holes given in Eqs.~(1)--(3) above.

Various models have been proposed in order to extract the rotational energy of black holes by processes of relativistic magnetohydrodynamics (see e.g., Ruffini and Wilson (1975)\cite{rw75}). It should be expected, however, that these processes are relevant over the long time scales characteristic of accretion processes.

In the present case of gamma ray bursts a sudden mechanism appears to be at work on time scales of the order of few seconds or shorter and they are naturally explained by the vacuum polarization process introduced in Damour \& Ruffini (1975)\cite{dr75}.

The fundamental new points we have found re-examining our previous work can be simply summarized, see Preparata, Ruffini and Xue (1998a,b)\cite{prx98ab} for details:

\begin{itemize}
\item The vacuum polarization process can  occur in an extended region around the black hole called the dyadosphere, extending from the horizon radius $r_+$ out to the dyadosphere radius $r_{ds}$. Only black holes with a mass larger than the upper limit of a neutron star and up to a maximum mass of $7.2\cdot 10^{6}M_\odot$ can have a dyadosphere.
\item The efficiency of transforming the mass-energy of a black hole into particle-antiparticle pairs outside the horizon can approach 100\%, for black holes in the above mass range. 
\item The created pairs are mainly positron-electron pairs and their number is much larger than the quantity $Q/e$ one would have naively expected on the grounds of qualitative considerations. It is actually given by $N_{\rm pairs}\sim{Q\over e}{r_{ds}\over \hbar/mc}$, where $m$ and $e$ are respectively  the electron mass and charge.  The energy of the pairs and consequently the  emission of the associated electromagnetic radiation  as a function of the black hole mass peaks in the gamma X-ray region.
\end{itemize}

Let us now recall the main results on the dyadosphere obtained in Preparata, Ruffini and Xue (1998a,b)\cite{prx98ab}. Although the general considerations presented by Damour and Ruffini (1975)\cite{dr75} refer to a rotating Kerr-Newmann field with axial symmetry about the rotation axis, for simplicity, we there considered the case of a nonrotating Reissner-Nordstr\"{o}m EMBH to illustrate the basic gravitational-electrodynamical process. The dyadosphere then lies between the radius 
\begin{equation}
r_{\rm ds}=\left({\hbar\over mc}\right)^{1\over2} \left({GM\over c^2}\right)^{1\over2} \left({m_{\rm p}\over m}\right)^{1\over2} \left({e\over q_{\rm p}}\right)^{1\over2} \left({Q\over\sqrt{G} M}\right)^{1\over2}
\label{rc}
\end{equation} 
and the horizon radius 
\begin{equation}
r_{+}={GM\over c^2}\left[1+\sqrt{1-{Q^2\over GM^2}}\right].
\label{r+}
\end{equation}
The number density of pairs created in the dyadosphere is 
\begin{equation}
N_{e^+e^-}\simeq {Q-Q_c\over e}\left[1+{(r_{ds}-r_+)\over {\hbar\over mc}}\right] \ ,
\label{n}
\end{equation}
where $Q_c=4\pi r_+^2{m^2c^3\over \hbar e}$. The total energy of pairs, converted from the static electric energy, deposited within the dyadosphere is then
\begin{equation}
E^{\rm tot}_{e^+e^-}={1\over2}{Q^2\over r_+}(1-{r_+\over r_{\rm ds}})(1-\left({r_+\over r_{\rm ds}}\right)^2) \,.
\label{tee}
\end{equation}

The analogies between the ergosphere and the dyadosphere are many and extremely attractive:

\begin{itemize}
\item Both of them are extended regions around the black hole.
\item In both regions the energy of the black hole can be extracted, approaching the limiting case of reversibility as from Christodoulou \& Ruffini (1971)\cite{ruffc}.
\item The electromagnetic energy extraction by the pair creation process in the dyadosphere is much simpler and less contrived than the corresponding process of rotational energy extraction from the ergosphere.
\end{itemize}

\section{the EM pulse of an atomic explosion versus the PEM pulse of a black hole}

The analysis of the radially resolved evolution of the energy deposited within the $e^+e^-$-pair and photon plasma fluid created in the dyadosphere of an EMBH is much more complex then we had initially anticipated. The collaboration with Jim Wilson and his group at Livermore Radiation Laboratory has been very important to us. We decided to join forces and propose a new collaboration with the Livermore group renewing the successful collaboration with Jim in 1974 (Ruffini and Wilson 1975)\cite{rw75}. We proceeded in parallel: in Rome with simple almost analytic models to be then validated by the Livermore codes (Wilson, Salmonson and Mathews 1997,1998)\cite{wsm97,wsm98}.

For the evolution we assumed the relativistic hydrodynamic equations, for details see Ruffini et al. (1998,1999)\cite{rswx98,rswx99}. We assumed the plasma fluid of $e^+e^-$-pairs, photons and baryons to be a simple perfect fluid in the curved space-time. The baryon-number and energy-momentum conservation laws are 
\begin{eqnarray}
(n_B U^\mu)_{;\mu}&=&(n_BU^t)_{,t}+{1\over r^2}(r^2 n_BU^r)_{,r}= 0\,, \label{contin}\\
(T_\mu^\sigma)_{;\sigma}&=&0\,,
\label{contine}
\end{eqnarray}
and the rate equation: 
\begin{equation}
(n_{e^\pm}U^\mu)_{;\mu}=\overline{\sigma v} \left[n_{e^-}(T)n_{e^+}(T) - n_{e^-}n_{e^+}\right] \,,
\label{econtin}
\end{equation}
where $U^\mu$ is the four-velocity of the plasma fluid, $n_B$ the proper baryon-number density, $n_{e^\pm}$ are the proper densities of electrons and positrons ($e^\pm$), $\sigma$ is the mean pair annihilation-creation cross-section, $v$ is the thermal velocity of the $e^\pm$, and $n_{e^\pm}(T)$ are the proper number-densities of the $e^\pm$ at an appropriate equilibrium temperature $T$. The calculations are continued until the plasma fluid expands, cools and the $e^+e^-$ pairs recombine and the system becomes optically thin.

The results of the Livermore computer code are compared and contrasted with three almost analytical models: (i) spherical model: the radial component of the four-velocity is of the form $U(r)=U{r\over {\cal R}}$, where $U$ is the four-velocity at the surface ($r={\cal R}$) of the plasma, similar to a portion of a Friedmann model, (ii) slab 1: $U(r)=U_r={\rm const.}$, an expanding slab  with constant width ${\cal D}= R_\circ$ in the coordinate frame in which the plasma is moving, (iii) slab 2: an expanding slab with constant width $R_2-R_1=R_\circ$ in the comoving frame of the plasma. We compute the relativistic Lorentz gamma factor $\gamma$ of the expanding $e^+e^-$ pair and photon plasma.

\begin{figure}
\vspace{-.5cm}
\epsfxsize=\hsize
\begin{center}
\mbox{\epsfbox{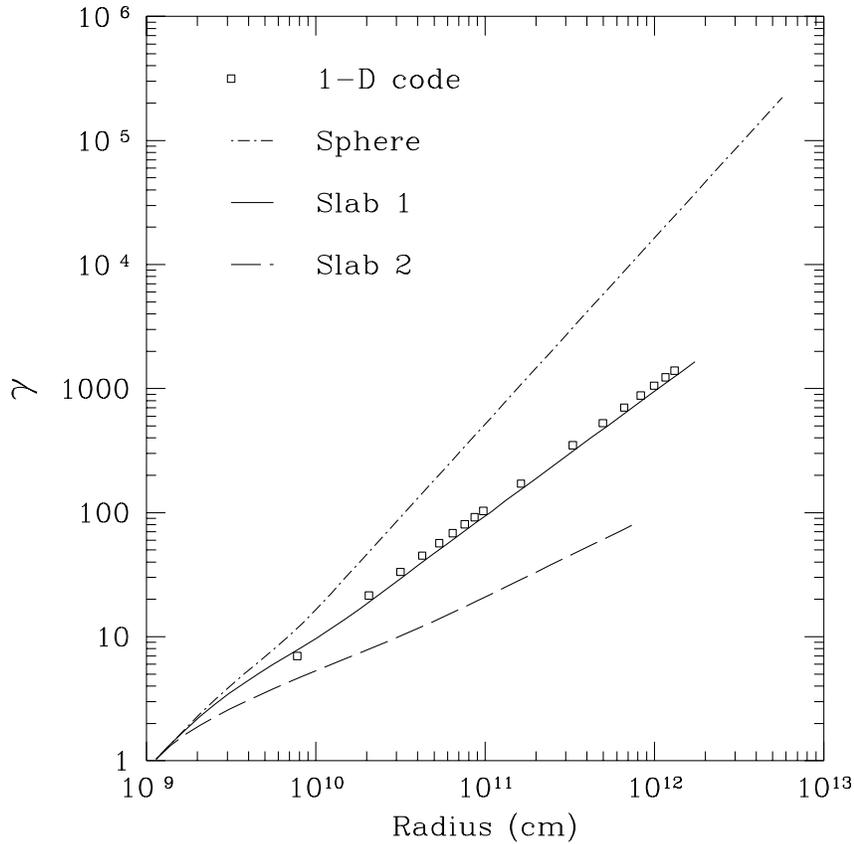}}
\end{center}
\vspace{-0.2cm}
\caption[]{Lorentz $\gamma$ as a function of radius. Three models for the expansion pattern of the $e^+e^-$ pair plasma are compared with the results of the one dimensional hydrodynamic code for a $1000 M_\odot$ black hole with charge $Q = 0.1 Q_{max}$.  The 1-D code has an expansion pattern that strongly resembles that of a shell with constant coordinate thickness. Reproduced from Ruffini, et al. (1999)\cite{rswx99}.}
\label{pic2}
\end{figure}

Figure (\ref{pic2}) shows a comparison of the Lorentz factor of the expanding fluid as a function of radius for all the models. One sees that the one-dimensional code (only a few significant points are plotted) matches the expansion pattern of a shell of constant coordinate thickness.

In analogy with the notorious electromagnetic radiation EM pulse of certain explosive events, we called this relativistic counterpart of an expanding pair electromagnetic radiation shell a PEM pulse.

\begin{figure}
\vspace{-.5cm}
\epsfxsize=\hsize
\begin{center}
\mbox{\epsfbox{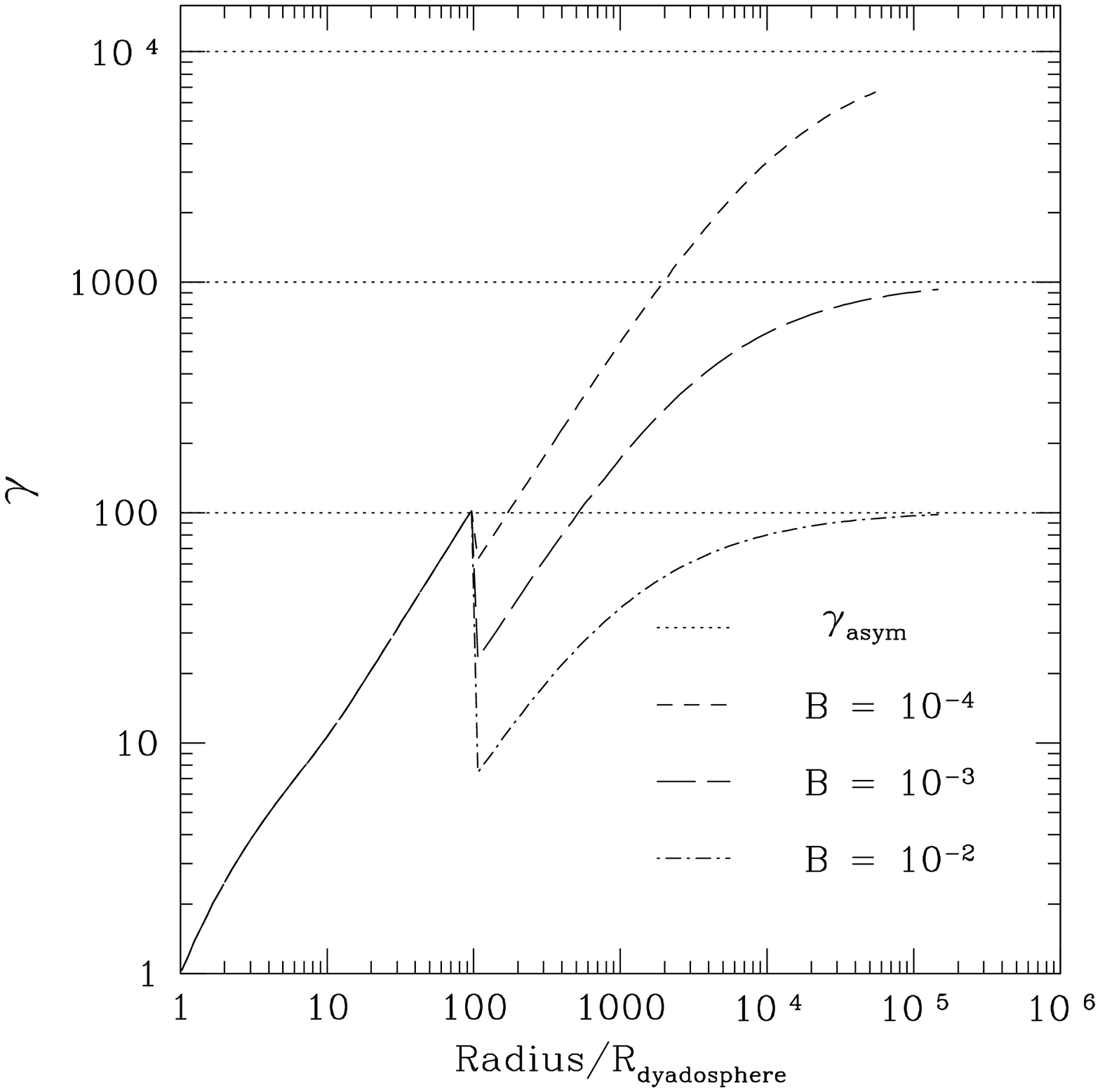}}
\end{center}
\vspace{-0.2cm}
\caption[]{Lorentz gamma factor $\gamma$ as a function of radius for the PEM pulse interacting with the baryonic matter of the remnant (PEMB pulse) for selected values of the baryonic matter. Reproduced from Ruffini, et al. (2000)\cite{rswx00}.}
\label{gammab}
\end{figure}

\begin{figure}
\vspace{-.5cm}
\epsfxsize=\hsize
\begin{center}
\mbox{\epsfbox{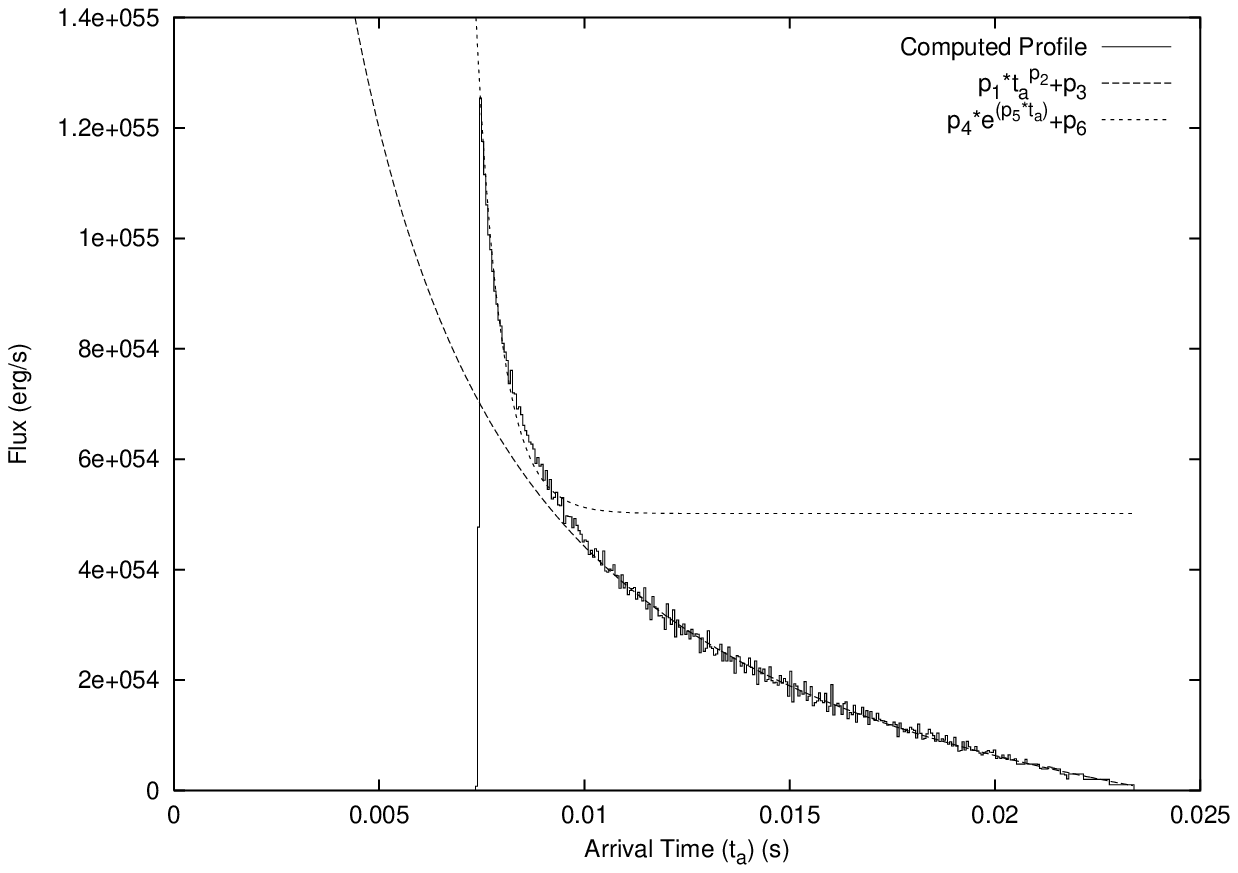}}
\end{center}
\vspace{-0.2cm}
\caption[]{P-GRB from an EMBH with $M=100 M_\odot$ and $Q = 0.1 Q_{max}$. Reproduced from Bianco, et al. (2001)\cite{brx01}.}
\label{pic3}
\end{figure}

In recent work we have computed the interaction of the expanding plasma with the surrounding baryonic matter (Ruffini, et al. 2000)\cite{rswx00}, see Fig~.\ref{gammab}. We have also been able to follow the expansion process all the way to the point where the transparency condition is reached and what we have defined the ``proper GRB'' (P-GRB) is emitted (Bianco, et al. 2001)\cite{brx01}, see Fig.~\ref{pic3}. We have then proceeded to develop the basic work to describe the afterglow of GRBs (Ruffini, et al. 2001)\cite{zzz01}. These results of our theoretical model have reached the point where they can be subjected to a direct comparison with the observational data.

\section{Three new paradigms for the interpretation of GRBs. Obtained after the Capri meeting}

Starting from this theoretical background, presented in CAPRI, we have moved ahead to fit the observational data on the basis of the EMBH model. We have used the GRB~991216 as a prototype, both for its very high energetics, which we have estimated in the range of $E_{\rm dya}\sim 9.57\times 10^{52}$ ergs, as well as for the superb data obtained by the Chandra and RXTE satellites. In order to understand the GRB phenomenon, we have found it necessary to formulate three new paradigms in our novel approach:

\begin{enumerate}
\item The Relative Space-Time Transformation (RSTT) paradigm. See Ruffini, Bianco, Chardonnet, Fraschetti, Xue (2001a)\cite{lett1}.
\item The Interpretation of the Burst Structure (IBS) paradigm. See Ruffini, Bianco, Chardonnet, Fraschetti, Xue (2001b)\cite{lett2}.
\item The Multiple-Collapse Time Sequence (MCTS) paradigm. See Ruffini, Bianco, Chardonnet, Fraschetti, Xue (2001c)\cite{lett3}.
\end{enumerate}

These results are currently being expanded in a detailed presentation.

\section{Conclusions}

In view of the above experiences, I have formulated some conclusions which may be of more general validity:

\begin{itemize}

\item Analogies. The analogies between classical regimes and general relativistic regimes have been at times helpful in giving the opportunity to glance on the enormous richness of the new physical processes contained in Einstein's theory of spacetime structure. In some cases they have allowed us to extend our knowledge and formalize new physical laws, the derivation of Eqs.~(1)--(3) is a good example. Such analogies have also provided dramatic evidence of the enormous differences in depth and physical complexity between classical physics and general relativistic effects. The cases of extracting rotational energy from a neutron star and from a rotating black hole are good examples.

\item New paradigms and their verifications. The establishment of new paradigms is essential to scientific process and is certainly not easy to do. Such paradigms are important in order to guide a meaningful comparison between theories and observations and much attention should be given to their development and inner conceptual consistency.Always the major factor driving the progress of scientific knowledge is the confrontation of theoretical predictions and the new paradigms of interpretation with observational data. In recent years the evolution of new technologies has permitted dramatic improvement in the sensitivity of the observational apparata. It is very gratifying that in this process of learning the structure of our Universe, the observational data intervene not in a marginal way, but with clear and unequivocal results: they confirm  the correct theories and their paradigms by impressive agreement and they disprove the wrong ones by equally impressive disagreement.

\item
The Gamma Ray astrophysics present a new area of research which transcends all preceeding ones. It implies the verification of general relativistic effects in totally unexplored and new regimes, it implies as well the extrapolation of current knowledge of subnuclear high energy physics into regimes again unexplored up to now on our planet, it needs technical developments in instrumentation both from space and from the ground unprecedented both for complexity, accuracy and need of coordination. The undestanding of this phenomena requires a totally novel style of research and is extremely promising for promoting the discovery of new fundamental physical laws in Nature.  

\end{itemize}

\end{document}